\documentclass[conference]{IEEEtran}
\IEEEoverridecommandlockouts
\usepackage[letterpaper, top=0.75in, bottom=1.05in, left=0.625in, right=0.625in]{geometry}
\usepackage{cite}
\usepackage{url}
\usepackage{amsmath,amssymb,amssymb}
\usepackage{algorithmic}
\usepackage{graphicx}
\usepackage{textcomp}
\usepackage{xcolor}
\usepackage{listings}
\usepackage{multirow} 
\usepackage{array}
\usepackage{booktabs}
\usepackage[ruled,vlined]{algorithm2e}
\usepackage{subcaption}
\usepackage{amssymb}
\usepackage{makecell} % 添加在文档导言区
\usepackage{pifont}
\usepackage{graphicx}
\newcommand{\xmark}{\ding{55}} % cross mark
\def\BibTeX{{\rm B\kern-.05em{\sc i\kern-.025em b}\kern-.08em
    T\kern-.1667em\lower.7ex\hbox{E}\kern-.125emX}}

\DeclareRobustCommand*{\IEEEauthorrefmark}[1]{%
	\raisebox{0pt}[0pt][0pt]{\textsuperscript{\footnotesize\ensuremath{#1}}}}

\usepackage[noabbrev,nameinlink]{cleveref}

\Crefformat{figure}{#2Fig.~#1#3}
\Crefmultiformat{figure}{Figs.~#2#1#3}{ and~#2#1#3}{, #2#1#3}{ and~#2#1#3}

\begin{document}

% \title{POLARIS: A Flexible Cross-Domain Access Control Architecture Driven by Self-Sovereign Identity}

\title{POLARIS: Cross-Domain Access Control via Verifiable Identity and Policy-Based Authorization}

\author{
	\IEEEauthorblockN{Aiyao Zhang\IEEEauthorrefmark{1,2}, Xiaodong Lee\IEEEauthorrefmark{1,3,4}\textsuperscript{,\textsection}, Zhixian Zhuang\IEEEauthorrefmark{1,2}, Jiuqi Wei\IEEEauthorrefmark{1,2}, Yufan Fu\IEEEauthorrefmark{1,2}, Botao Peng\IEEEauthorrefmark{1,3}}
	
	\IEEEauthorblockA{\IEEEauthorrefmark{1}Institute of Computing Technology, Chinese Academy of Sciences}
	\IEEEauthorblockA{\IEEEauthorrefmark{2}University of Chinese Academy of Sciences}
        \IEEEauthorblockA{\IEEEauthorrefmark{3}Fuxi Institution}
        \IEEEauthorblockA{\IEEEauthorrefmark{4}Center for Internet Governance, Tsinghua University}
	\IEEEauthorblockA{\{zhangaiyao22z, xl, zhuangzhixian22s, weijiuqi19z, fuyufan20z, pengbotao\}@ict.ac.cn}
} 

\maketitle

\begingroup\renewcommand\thefootnote{\textsection}
\footnotetext{Xiaodong Lee is the corresponding author.}
\endgroup

\begin{abstract}
Access control is a security mechanism designed to ensure that only authorized users can access specific resources. Cross-domain access control involves access to resources across different organizations, institutions, or applications. Traditional access control, however, which handles authentication and authorization separately in centralized environments, faces challenges in identity dispersion, privacy leakage, and diversified permission requirements, failing to adapt to cross-domain scenarios. Thus, there is an urgent need for a new access control mechanism that empowers autonomous control over user identity and resources, addressing the demands for privacy-preserving authentication and flexible authorization in cross-domain scenarios. 

To address cross-domain access control challenges, we propose POLARIS, a unified and extensible architecture that enables policy-based, verifiable and privacy-preserving access control across different domains. POLARIS features a structured commitment mechanism for reliable, fine-grained, policy-based identity disclosure. It further introduces VPPL, a lightweight policy language that supports issuer-bound evaluation of selectively revealed attributes. A dedicated session-level security mechanism ensures binding between authentication and access, enhancing confidentiality and resilience to replay attacks.

We implement a working prototype and conduct comprehensive experiments, demonstrating that POLARIS effectively provides scalable, privacy-preserving, and interoperable access control across heterogeneous domains. Our results highlight the practical viability of POLARIS for enabling secure and privacy-preserving access control in decentralized, cross-domain environments.
\end{abstract}


\section{Introduction}

\subsection{Motivation}
Access control governs the enforcement of permissions to ensure that only authorized users can access specific resources\cite{intro_0}. It typically involves verifying user identity and managing access to data and system resources based on predefined permission policies\cite{intro_1}. Cross-domain access control refers to the implementation of access control mechanisms across different systems, organizations, platforms, or networks, where domains are inherently untrusted and lack direct interoperability\cite{intro_2}. Nowadays, an increasing range of scenarios involve cross-domain access control\cite{intro_4, dia, intro_5}. For example, in a supply chain, participants such as suppliers, manufacturers and logistics companies need to ensure data access is restricted to what is necessary for their operations\cite{intro_5}. Similarly, in the healthcare sector, institutions need to limit data access to specific, authorized entities to safeguard sensitive patient information\cite{smartaccess}.

Traditional access control mechanism generally involves two steps: authentication, which verifies the authenticity of the user's identity, and authorization, which determines access rights to resources\cite{intro_7}. This mechanism is well-suited for single domain, where user information is centrally stored, enabling straightforward authentication and authorization. However, cross-domain access control introduces new challenges for both authentication and authorization: Identities are no longer stored in a single domain, but are managed by users in a decentralized environment\cite{ssi_survey}. This involves complex, heterogeneous identity and varying authentication methods, while users typically prefer to share only the necessary information with identity verifier, rather than disclosing excessive personal data, raising privacy and security issues. These issues also bring new challenges to authorization. Traditional authorization methods (e.g., ACLs, RBAC) are primarily based on static and centralized models, where predefined permissions are assigned to specific users or roles, limiting their adaptability in dynamic and complex cross-domain scenarios\cite{intro_8}. 

In summary, traditional access control mechanism cannot effectively adapt to cross-domain scenarios, a new architecture is urgently needed to solve the above challenges: A decentralized authentication framework is required to ensure the credibility and autonomy of identities across heterogeneous ecosystems, while a verifiable and expressive access control model must provide fine-grained control anchored in trusted identity sources and enriched with semantic policy expressiveness, ensuring trustworthy and privacy-preserving authorization.

\subsection{Our Solution}
We propose POLARIS (\textbf{POL}icy-driven \textbf{A}ccess control with \textbf{R}eliable \textbf{I}dentity \textbf{S}upport), a unified and extensible architecture that enables policy-based, verifiable and privacy-preserving access control across different domains. POLARIS establishes a secure cross-domain authentication architecture based on decentralized identifiers~\cite{did}, and introduces a novel Structured Commitment Disclosure (SCD) mechanism to enable privacy-preserving authentication by minimizing personal data exposure while ensuring data integrity. Building on this authentication scheme, we propose the Verifiable Presentation Policy Language (VPPL), a lightweight and verifiable policy language that enables fine-grained access control and trustworthy policy expression over selectively disclosed claims, thereby enabling an end-to-end decentralized authorization workflow.

Our contributions are summarized as follows:
\begin{itemize}
    \item We propose \textbf{POLARIS}, a unified and extensible architecture for cross-domain access control, which integrates policy-based authorization and verifiable credential presentation in self-sovereign identity (SSI) environments.
    
    \item We design a \textit{Structured Commitment Disclosure (SCD)} mechanism that enables users to selectively disclose attributes from verifiable credentials while ensuring integrity and verifiability, thereby achieving reliable and privacy-enhanced decentralized identity verification.

    \item We introduce \textit{VPPL}, a lightweight and verifiable policy expression language that allows resource owners to flexibly define fine-grained access control policies over selectively revealed attributes with issuer-level trust anchoring.

    \item We develop a \textit{session-level security mechanism} based on key derivation, which binds authentication with resource access, enhancing session confidentiality, integrity, and replay resistance.

    \item We implement a prototype of POLARIS and conduct comprehensive experiments and comparative analysis, demonstrating its high efficiency, concurrency performance, and architectural advantages over representative state-of-the-art approaches.
\end{itemize}

\begin{figure*}[!t]
\centering
\includegraphics[width=0.92\textwidth]{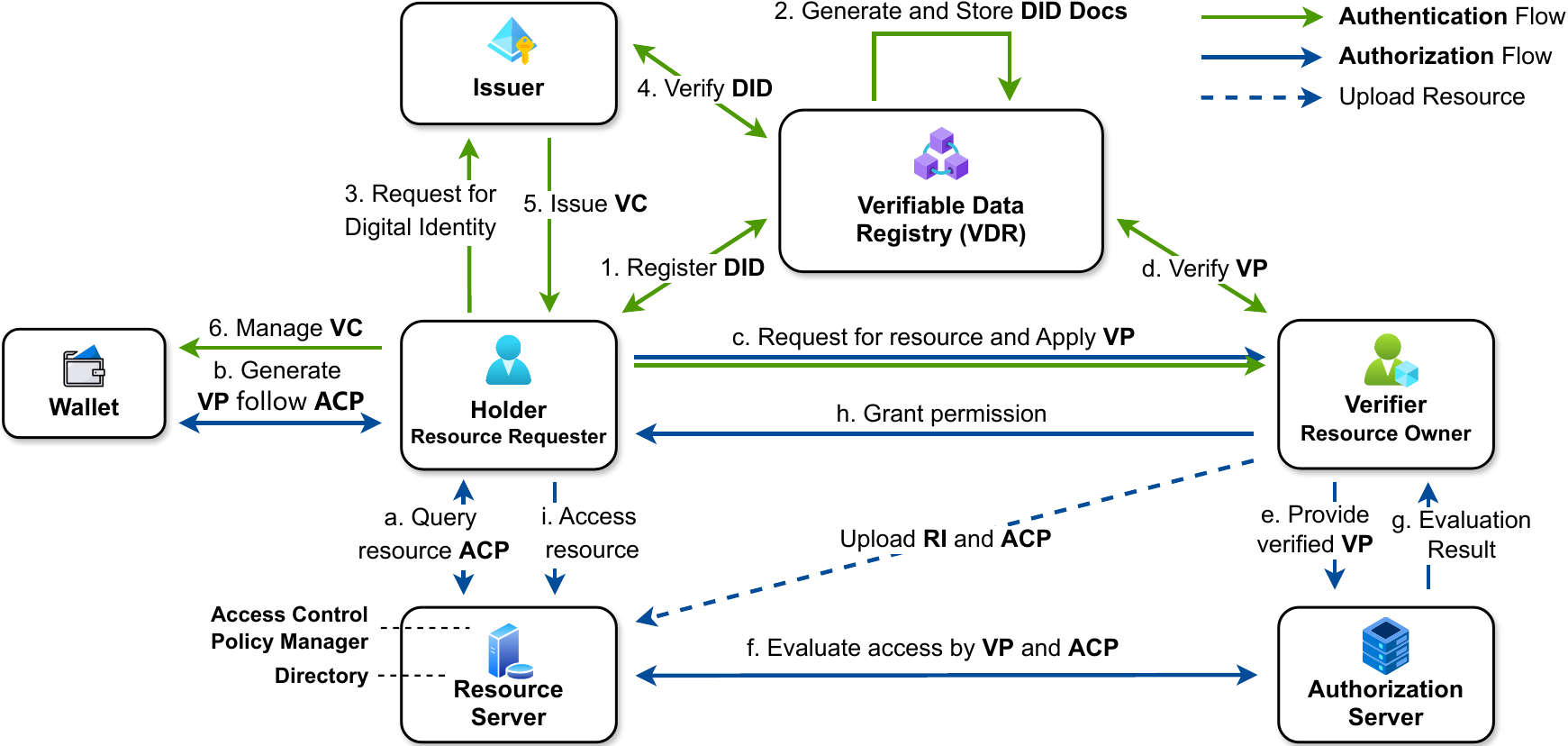}
\caption{Components and interaction flow of POLARIS.}
\label{fig:sifaa}
\vspace{-5pt}
\end{figure*}

\section{Related Work}\label{related_work}
As information systems evolve toward multi-party collaboration and cross-domain data sharing, access control faces critical challenges such as identity heterogeneity, trust decoupling, and privacy leakage. Existing research has explored privacy-enhancing technologies, protocol specification, and architecture integration, aiming to improve selective disclosure capabilities, support decentralized identity authentication, and establish scalable access control mechanisms.

\textbf{At the technological level}, BBS+\cite{bbs+} offers a signature scheme that supports selective disclosure and unlinkability, enabling multi-attribute privacy while preserving verifiability. However, it remains computationally heavy and difficult to scale in resource-constrained environments. Other works~\cite{zkp1,zkp2,zkp3} construct anonymous credentials and range proofs using zero-knowledge proofs (ZKPs), achieving stronger privacy guarantees but still facing limitations in versatility, scalability, and deployment efficiency.

\textbf{At the specification level}, the development of decentralized identity systems has promoted the standardization of core specifications such as decentralized identifiers (DIDs)\cite{did} and verifiable credentials (VCs)\cite{vc} by W3C, laying the foundation for identity interoperability and credential exchange in decentralized settings. Building upon these, OIDC4VC\cite{oidc4vci} and OIDC4VP\cite{oidc4vp} introduced VC capabilities into the OpenID Connect process to support the acquisition and presentation of verifiable credentials. However, they still rely on centralized identity providers and trust anchors, limiting their adaptability to decentralized collaboration scenarios. Meanwhile, the IETF draft SD-JWT\cite{sd-jwt} proposes a selective disclosure mechanism for attributes based on hash summaries, which achieves minimum disclosure at the declaration level without introducing ZKPs, lowering the implementation threshold of privacy enhancement mechanisms. However, it lacks support for policy expression and access control processes, focusing primarily on the identity representation layer.

\textbf{At the architectural level}, studies such as Lagutin~\cite{Lagutin} and Jung~\cite{Jung} focus on lightweight integration of SSI/DID into traditional models such as OAuth2 or RBAC, improving identity portability and decentralization to some extent but lacking support for traceable auditing and fine-grained policy execution. Other approaches~\cite{ssibac, Ma, Pandiyarajan} integrate verifiable credentials and zero-knowledge proofs into ABAC-based architectures, enhancing expressiveness and security. However, such solutions usually have high policy execution complexity, and the traditional policy language is structurally incompatible with the JSON model of VC/VP, which often require additional conversion and adaptation logic in actual deployment. Approaches like \cite{Saidi, flexauth, Kamboj} focus on on-chain public key-based authorization mechanisms, but pay limited attention to user-controllable disclosure capabilities and collaborative binding of multi-source credentials.

Overall, existing research primarily addresses isolated aspects such as identity representation or privacy disclosure, while lacking an end-to-end framework that seamlessly supports identity registration, credential issuance, policy enforcement, and access verification in a unified manner.

\section{Design of POLARIS} \label{system_design}
In this section, we describe the design of POLARIS, a unified and extensible architecture that supports verifiable, policy-based, and privacy-preserving access control across domains. We begin by outlining the system components, followed by detailed descriptions of SSI identity management and VP-based access control process. Lastly, we describe a session-level security mechanism designed to safeguard sensitive interactions across authentication and authorization phases.

\subsection{System Components} \label{4.1}
As shown in Figure.~\ref{fig:sifaa}, our system comprises the following entities and modules: Issuer, Holder (Resource Requester), Wallet, Verifier (Resource Owner), Verifiable Data Registry, Authorization Server and Resource Server. The specific functions of these components are described as follows:
\begin{itemize}
    \item \textbf{Issuer:} An authoritative entity that generates and signs \textbf{verifiable credentials (VCs)} to certify the Holder's digital identity, which may represent an individual, organization or institution.
    \item \textbf{Holder/Resource Requester:} In the process of access control, Holder also acts as Resource Requester. They use VCs provided by Issuer to prove their identity and interact with Resource Owner to gain access to the resources.
    \item \textbf{Wallet (End-User Component):} Each Holder manages their identity credentials and secret keys through their own wallet, generating a \textbf{verifiable presentation (VP)} through selective disclosure based on access control needs.
    \item \textbf{Verifier/Resource Owner:} Verifier also acts as Resource Owner, which is responsible for authenticating Resource Requester. Resource owners have absolute control over their resources. They can define Access Control Policies (ACPs) by interacting with Resource Server and grant permissions based on the policy evaluation results of Authorization Server.

    \item \textbf{Verifiable Data Registry (VDR):} Implemented through distributed platforms like blockchain, serving as foundational infrastructure for trust establishment. VDR provides secure and tamper-proof storage for DIDs and their associated documents, and supports key operations such as identifier resolution and public key verification essential for authentication.

    \item \textbf{Authorization Server (AS)}: Comprising a Policy Decision Point (PDP) and a Policy Enforcement Point (PEP). The PDP receives the VP from the Resource Requester and retrieves the ACP from the Resource Server to perform verifiable policy evaluation. PEP returns the policy evaluation results to the resource owner, who then determines how to grant access permissions to the requester according to their preferences.
    \item \textbf{Resource Server (RS):} Comprising an \textbf{Access Control Policy Manager (ACPM)} and \textbf{Directory}. ACPM manages ACPs for resources. Directory maintains a resource directory, which contains basic information of resources and helps locate the specific storage point. This storage point can be located on a cloud server, a local edge gateway, or a personal device, depending on the resource owner.
\end{itemize}

\subsection{Self-Sovereign Identity Management} \label{4.2}

The Green Arrow in Figure.~\ref{fig:sifaa} is the authentication part of cross-domain access control. It follows the DID and VC specifications in W3C and supports the reliable issuance and verification of identity credentials from different domains. We will introduce the whole process by two parts: digital identity acquisition flow and selective disclosure mechanism.

\subsubsection{\textbf{Digital Identity Acquisition Flow}} 
\begin{figure}[h]
\centering
\includegraphics[width=0.48\textwidth]{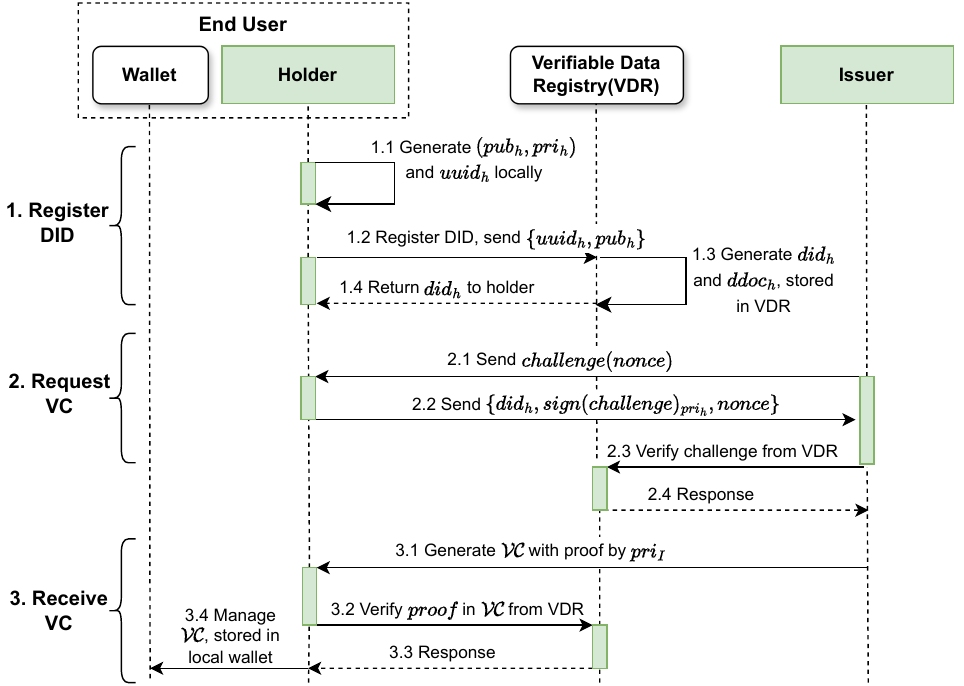}
\caption{Digital Identity Acquisition Flow for Decentralized Registration and Credential Issuance.}
\label{fig:did}
\vspace{-5pt}
\end{figure}

Figure.~\ref{fig:did} shows the process of digital identity acquisition, which mainly includes the steps of register DID, request VC and receive VC.

\textbf{Register DID.} To register a DID in POLARIS, Holder first generates a local asymmetric key pair $\{pub_h, pri_h\}$ and a unique identifier $uuid_h$ (step 1.1), then submits them to VDR (step 1.2). Upon receiving the request, VDR issues a globally unique decentralized identifier $did_h$, constructs the corresponding DID document $ddoc_h$ containing $pub_h$, and stores it in VDR (step 1.3). Finally, VDR returns $did_h$ to Holder to complete the registration (step 1.4).

\textbf{Request VC.} After registration, Holder requests a digital identity credential from Issuer by proving possession of the private key associated with $did_h$ through a challenge–response process. Specifically, Issuer sends a fresh challenge containing a nonce to prevent replay attacks (step 2.1), and Holder signs the challenge using $pri_h$ to demonstrate key ownership (step 2.2). Issuer then resolves $did_h$ via the VDR to obtain $pub_h$ and verifies the signature and nonce validity (step 2.3 and 2.4).

\textbf{Receive VC.} Once the Holder's identity is verified, Issuer signs and provides VC to the Holder (step 3.1). Holder should first verify Issuer's authenticity by checking $pub_I$ in VDR (step 3.2 and 3.3). After successful verification, VC is stored and managed in Holder's digital wallet (step 3.4).

\subsubsection{\textbf{Selective Disclosure Mechanism}}\label{selective_disclosure}
Traditional VC consists of \textbf{Metadata} (basic properties like issuer and validity date), \textbf{Claims} (identity information claimed by Issuer) and \textbf{Proof} (signature of Issuer to Metadata and Claims). However, this structure can raise privacy concerns, as Holder may only need to disclose specific claims from the VC, without revealing full personal details. To address this, we introduce the Structured Commitment Disclosure (SCD) mechanism, to enable privacy-preserving and selective disclosure of identity.

SCD mechanism generates a random salt $s_i$ for each attribute, concatenates it with the attribute name $A_i$ and value $v_i$, and produces a cryptographic commitment for the attribute by hashing the concatenated string, as illustrated in Equation~\eqref{hClaims}. All hashed claims are organized as a structured map indexed by attribute names.
% Here, $A_i$, $v_i$, and $s_i$ denote the attribute name, value and random salt respectively. 
The inclusion of random salt mitigates the risk of enumerating low-entropy attributes (e.g., gender, education) through brute-force attempts.

\begin{equation}
    \begin{aligned}
        hClaims &= \{A_1: c_1,\; A_2: c_2,\; \dots,\; A_n: c_n\}, \\
        c_i     &= Hash(A_i \,\|\, v_i \,\|\, s_i)
    \end{aligned}\label{hClaims}
\end{equation}

When Issuer issues a digital identity certificate to Holder, it needs to provide the hashed Claims, together with the salt sequence used to generate the commitment. We use $\mathcal{VC}$ to present the verifiable credential, which contains $Metadata$, $hClaims$ and $\sigma$, as Equation~\eqref{VC}.

\begin{equation}
\begin{aligned}
    \mathcal{VC} &= \{Metadata, hClaims, \sigma\}\label{VC} \\
    \sigma = &Signature\{Metadata \| hClaims\}_{pri_{I}}
\end{aligned}
\end{equation}

$\mathcal{VC}$ can realize the authentication of any claimed attributes without exposing all attributes, satisfying the concealment and binding of the commitment scheme. To achieve selective disclosure, the Verifier only receives the attributes to be disclosed, denoted as $A_v = \{(A_i, v_i), ..., (A_j, v_j)\}$, along with the corresponding salt sequence $S = \{s_i, ..., s_j\}$ and the original $\mathcal{VC}$.

During authentication process, Holder can aggregate multiple $\mathcal{VC}$s issued by different trusted parties into a $\mathcal{VP}$, which includes selectively disclosed attributes and their structured commitments. The message submitted to the Verifier consists of $\mathcal{VP}$ and the Holder’s signature, as defined in Equation~\eqref{eq:vp}.

\begin{equation}
    \begin{aligned}
        &Message = \{\mathcal{VP}, \sigma'\} \\
        &\mathcal{VP} = \{\mathcal{VC}_k, Av_k, S_k\}_{k=1}^n \\
        &Av_k = \{(A_i^k, v_i^k), ..., (A_j^k, v_j^k)\}, S_k = \{s_i^k, ..., s_j^k\} \\
        &\sigma' = Signature\{\mathcal{VP}\}_{pri_H}
    \end{aligned}
    \label{eq:vp}
\end{equation}

Upon receiving the message, Verifier performs identity verification through a three-step process:
\begin{enumerate}
    \item \textbf{Holder authentication:} Verify the Holder’s signature $\sigma'$ in $Message$ to confirm the authenticity and integrity of the submission.
    \item \textbf{Issuer binding verification:} For each $\mathcal{VC}_k$ contained in $\mathcal{VP}$, validate its embedded proof $\sigma$ to ensure that the hashed claims and metadata were issued by a trusted authority, as:
    \[
         VerifySignature(\mathcal{VC}_k)=true,
    \]
    \item \textbf{Attribute-level verification:} For each disclosed attribute $(A_i, v_i)$ in $Av_k$ with corresponding salt values $s_i$, compute the hash $c_i' = Hash(A_i \| v_i \| s_i)$, and compare it against the corresponding entry in the structured commitment map, as:
    \[
        c_i' \stackrel{?}{=} \mathcal{VC}_k.\mathit{hClaims}[A_i]
    \]
\end{enumerate}

\begin{figure*}[htpb]
\centering
\includegraphics[width=0.92\textwidth]{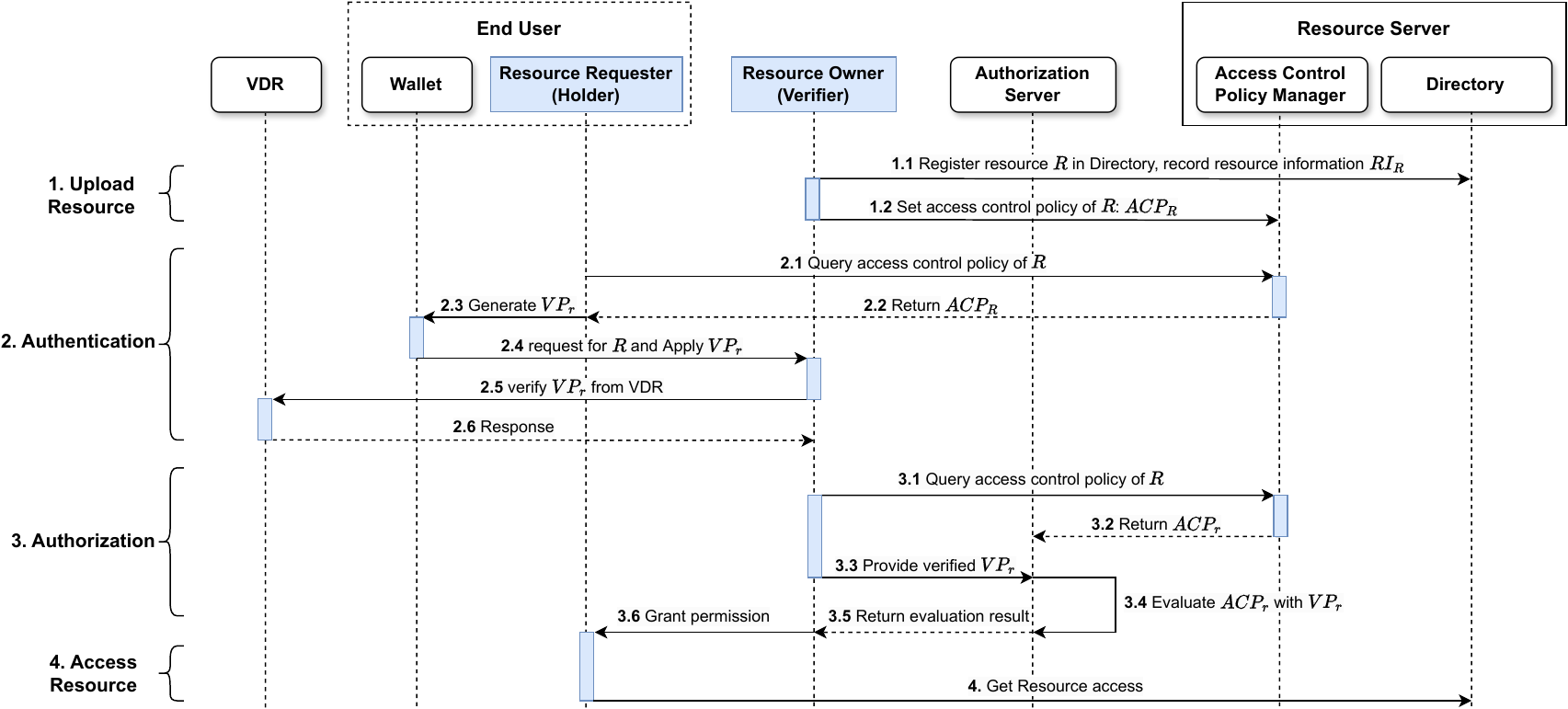}
\caption{Access Control Flow for Cross-Domain Authentication and Authorization.}
\label{fig:oidc}
\end{figure*}

The structured indexing of $hClaims$ enables direct attribute lookup, avoiding sequential search and supporting efficient verification across $\mathcal{VP}.$ This facilitates unified authentication of attribute claims from heterogeneous credentials in cross-domain settings. The verified attributes are naturally aligned with access control policy constraints, thereby enabling the integration of authentication and authorization, as introduced in Section~\ref{VPAC}.

\subsection{Verifiable Presentation-Based Access Control} \label{VPAC}
The blue arrow in Figure.~\ref{fig:sifaa} illustrates the authorization flow in cross-domain access control, which together with the decentralized authentication mechanism, forms a complete framework. To support these interactions, we propose a VP-based access control framework integrating selective disclosure, verifiable attribute matching, and policy-based authorization. This section presents the overall access control flow, introduces the policy expression language, and describes the policy evaluation process over VP submissions.

\subsubsection{\textbf{Access Control Flow}} \label{lab:overall_process}

Figure.~\ref{fig:oidc} shows the overall access control flow in POLARIS, consisting of upload resource, authentication, authorization and access resource.

\textbf{Upload Resource.} Initially, when the Resource Owner uploads a data resource $R$ in POLARIS, the metadata $RI_R$ is recorded in the Directory (step 1.1), which contains the resource's address and ownership information. Additionally, the corresponding Access Control Policy $ACP_R$ is generated by Resource Owner and stored in the ACPM (step 1.2).

\textbf{Authentication.} Resource Requester retrieves the intended $ACP_R$ from ACPM (step 2.1 and 2.2) and constructs $VP_r$ via its wallet (step 2.3) through SCD mechanism. The Requester submits $VP_r$ and sends a request for $R$ (step 2.4), after which Resource Owner verifies VP's authenticity and integrity (step 2.5 and 2.6) to complete authentication.

\textbf{Authorization.} After successful authentication, the Resource Owner queries $ACP_R$ and forwards the verified $VP_r$ to the Authorization Server (step 3.1 to 3.3). The Authorization Server evaluates $ACP_R$ against the claims in $VP_r$ (step 3.4), and returns evaluation result back to Resource Owner (step 3.5), who then makes the final access decision and grant the permission to Requester (step 3.6).

\textbf{Access Resource.} After receiving permission based on the evaluated policy, the Resource Requester actively accesses the protected resource $R$ through the authorization service (step 4).

\begin{figure*}[!t]
\centering
\includegraphics[width=0.95\textwidth]{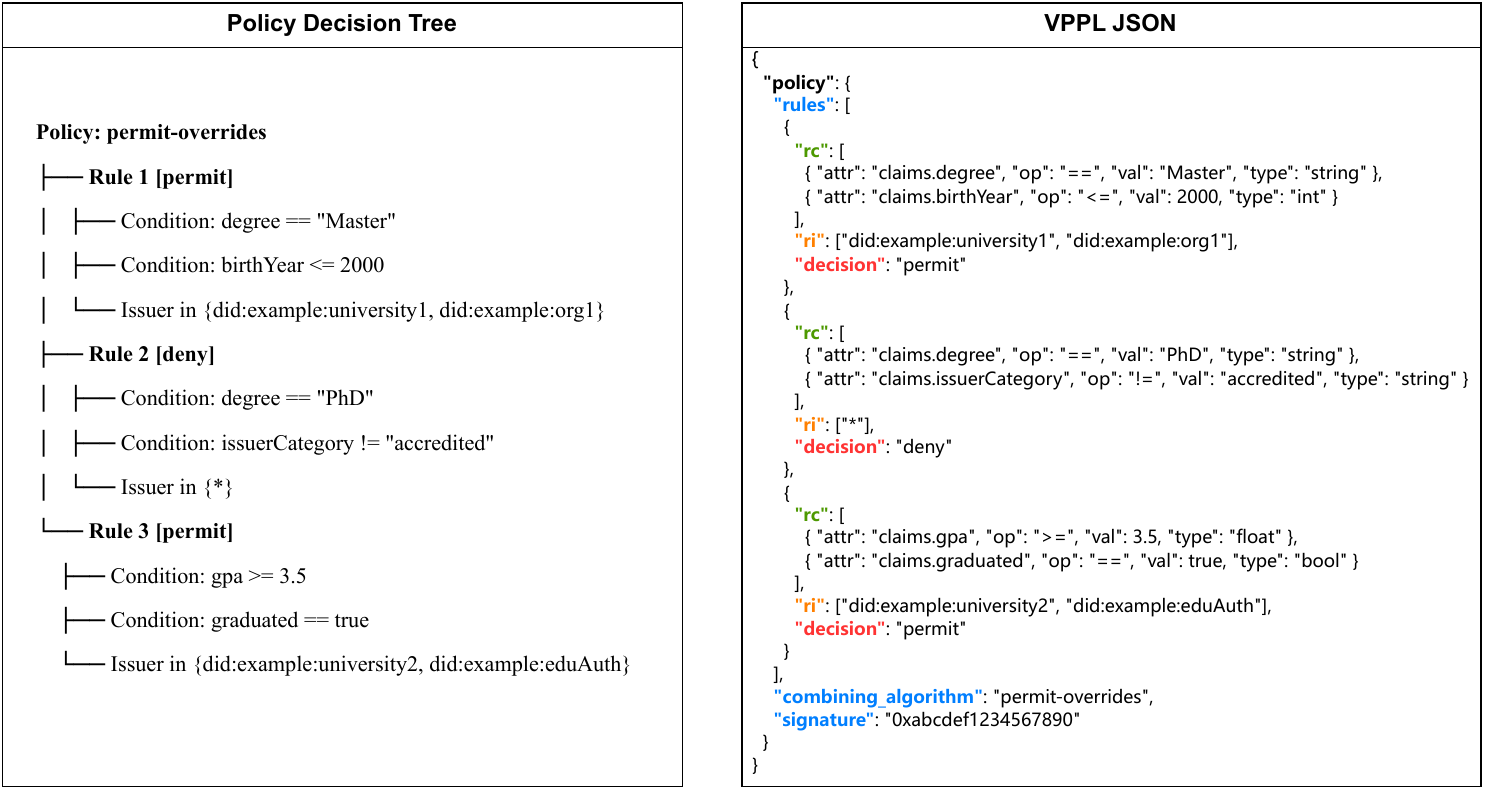}
\caption{VPPL Representation of Access Control Policy: Semantic Rule Tree and JSON Format}
\label{fig:acp}
\vspace{-5pt}
\end{figure*}

\subsubsection{\textbf{Policy Expression Language Design}}
In cross-domain access control, identity description and attribute definitions vary greatly between different domains, and access control policies require high scalability and customization capabilities. Although the traditional ABAC model\cite{abac} supports complex authorization requirements, it relies on a centralized identity source and lacks policy credibility, attribute verifiability, and support for selective disclosure.

To address these limitations, we propose the Verifiable Presentation Policy Language (VPPL), a policy expression language designed for verifiable identities. VPPL builds upon the ABAC model and semantically aligns with XACML's rule composition and design logic, while introducing a lightweight JSON-based syntax consistent with the standard data models of VCs and VPs. This design improves interoperability and deployment flexibility, thereby facilitating adoption in cross-domain environments. In addition, VPPL supports explicit binding between issuers and attributes within policy rules, enabling fine-grained control over trusted identity sources and strengthening the overall trust linkage between identity verification and policy enforcement.

Formally, a VPPL policy consists of multiple rules, a combining algorithm (e.g., \texttt{permit-overrides}), and an optional signature, as shown in Equation~\eqref{eq:policy}.

\begin{equation}
    Policy = (\{Rule\}, \texttt{Comb}, \texttt{Signature}) \label{eq:policy}
\end{equation}

The optional $\texttt{Signature}$ field ensures the authenticity and integrity. The resource owner may sign the serialized VPPL document, allowing the PDP to verify that it has not been altered prior to evaluation. This prevents unauthorized modification and establishes a verifiable trust linkage between resource owners and their access control polices.

Each authorization rule is defined as shown in Equation~\eqref{eq:rule}, where $R_C$ is a set of attribute-based expressions representing access conditions, $R_I$ specifies acceptable issuer DIDs, and $D \in \{\texttt{permit}, \texttt{deny}\}$ denotes the decision outcome.
\begin{equation}
    \begin{aligned}
        &Rule=(R_C, R_I, D)\label{eq:rule} \\
        &R_C = \{\text{Expr}_1, \text{Expr}_2, \dots, \text{Expr}_n\}
    \end{aligned}
\end{equation}

Each attribute-based expression is defined as Equation~\eqref{eq:expr}, where $A_i$ denotes the attribute path (e.g., \texttt{"claims.degree"}), $f_i$ represents the function (e.g., \texttt{==}, \texttt{in}, \texttt{>=}), $v_i$ and $t_i$ denotes the target value and data type (e.g., \texttt{string}, \texttt{int}, \texttt{bool}).

\begin{equation}
\text{Expr} = (A_i, f_i, v_i, t_i)\label{eq:expr}
\end{equation}

VPPL maintains atomic predicates at the expression level, while complex Boolean logic is achieved through rule-level composition and standard combining algorithms (e.g., \texttt{permit-overrides}), preserving expressiveness without introducing nested syntax or evaluation ambiguity.

For illustration, Figure.~\ref{fig:acp} presents a VPPL policy containing multiple rule conditions, issuer constraints and a permit-overrides combining algorithm. The policy is represented in both VPPL JSON syntax and an equivalent semantic rule tree, demonstrating how VPPL can express fine-grained, multi-issuer access decisions in a structured and interpretable form.

\subsubsection{\textbf{Policy Evaluation over VP}}
To evaluate a VP against a policy, the PDP parses the selectively disclosed claims and checks whether they satisfy all attribute-based expressions defined in each rule. As described in Section \ref{selective_disclosure}, $\mathcal{VP}$ is formalized as $\{\mathcal{VC}_k, Av_k, S_k\}_{k=1}^n$, where each $\mathcal{VC}$ presents a hashed verifiable credential issued by a distinct issuer, and $Av_k$ is the set of disclosed claims from that credential, along with corresponding salt set $S_k$.

\begin{equation}
    \begin{aligned}
        A_i = A_j, \quad& f_j(v_i,v_j^*) = true, \\
        Hash(A_i\|v_i\|s_i) &= \mathcal{VC}_k.hClaims[A_i], \\
        \mathcal{VC}_k.Meta&data.issuer \in R_I
    \end{aligned} \label{eq:vppl}
\end{equation}

To determine whether the rule $\mathcal{R} = (R_C, R_I, D)$ is satisfied, for every expression $(A_j, f_j, v_j^*, t_j) \in R_C$, there exists a verifiable credential $\mathcal{VC}_k$ whose claim $(A_i, v_i)$ satisfies the matching conditions. In addition, each $VC_k$ must be both cryptographically authentic and issued by the issuer specified by the authorization rule, as shown in Equation~\eqref{eq:vppl}.

Compared with traditional ABAC, our approach introduces three critical advancements that are non-trivial to achieve in decentralized settings:

\begin{itemize}
    \item \textbf{Cryptographic Verifiability.} VPPL binds each disclosed attribute to its issuer through hash commitments, enabling verifiable and tamper-resistant policy evaluation without relying on trusted attribute sources.
    \item \textbf{Privacy-Preserving Authorization.} Instead of exposing full identity information, VPPL supports fine-grained evaluation over selective disclosure. Verifier receives only the minimal information necessary to enforce the policy, preserving user privacy while maintaining policy correctness.
    \item \textbf{Decentralized Policy Integrity.} Each VPPL policy can be digitally signed and shared across domains, enabling verification of its origin and integrity. This ensures interoperable and trustworthy policy enforcement and facilitates integration with on-chain access systems.
\end{itemize}

VPPL extends the traditional XACML model to address the challenges posed by decentralized systems, particularly those based on VCs and VPs. It reconstructs the policy expression and evaluation process to support selectively disclosed attributes and claim-level issuer validation, enabling cryptographic enforcement of attribute constraints without relying on centralized identity providers. These enhancements collectively enable fine-grained, privacy-preserving, and interoperable access control in decentralized identity environments.

\subsection{Session-Level Security Mechanism}\label{key_derivation}
Typical processes such as identity acquisition and access control, inherently involve cross-entity data exchanges that transmit structured and sensitive information. These interactions require security capabilities including secure communication, identity binding, and fine-grained authorization to ensure confidentiality and compliance.

To this end, we design and implement a unified session-level key derivation mechanism, embedded in various key interaction paths as the system’s underlying data protection module. This mechanism is based on the standard PBKDF2 algorithm~\cite{pbkdf2}. The initiator automatically generates a symmetric key when each session is initialized, and synchronizes it after encrypting it with the other party’s public key. The derived key is only valid for the current session and expires after a timeout, thereby enhancing communication confidentiality, resistance to replay attacks, and session-level isolation.

Specifically, before participants $A$ and $B$ start interacting and transmitting message $M$, the initiator $A$ derives the session key $dKey$ through Equation~\eqref{eq:dkey}, where for automation of the entire process, we set the $password$ as $\{did_A \| did_B\}$ by default.
\begin{equation}
    dKey = PBKDF2(password, salt, keySize, iter) \label{eq:dkey}
\end{equation}

Once the key is generated, $A$ stores $dKey$ locally to handle all encryption, decryption, and signing operations for the current session. Subsequently, $A$ encrypts the derived key using PKI for transmission, enabling key synchronization between $A$ and $B$. Since PKI encryption and signature verification are performed only once per session, the overhead is manageable in the overall context. Finally, the message $M$ is encrypted using $dKey$. The derived-key encrypted message along with the public-key encrypted $dKey$, is sent as $\widetilde{M}$ to participant $B$, as Equation~\eqref{message} describes.
\begin{equation}
    \begin{aligned}
        &\widetilde{M} = \{DerivedKeyEncrypt(M, dKey),  \\
        &PublicKeyEncrypt(dKey, pub_B)\} \label{message}
    \end{aligned} 
\end{equation} 

This session-level security mechanism demonstrates the following three advantages in practice:
\begin{itemize}
    \item \textbf{Strong session isolation:} A unique session key is derived for each interaction, limiting the impact of key compromise to a single session and supporting fine-grained access control through scoped privilege enforcement.
    \item \textbf{Support for large data transmission:} Symmetric encryption overcomes the ciphertext size limitations of Public Key Infrastructure (PKI) cryptography, making it suitable for securely encapsulating structured credentials such as VPs and VCs.
    \item \textbf{Efficiency Improvement:} PKI encryption is performed only once during session initiation for key synchronization, while subsequent communications rely entirely on efficient symmetric encryption, eliminating the need for repeated on-chain public key retrievals and thereby reducing system latency and blockchain-related overhead.
\end{itemize}

This mechanism operates independently of global key management and is applicable across diverse authentication and authorization scenarios, significantly enhancing the system’s security resilience and permission control in cross-domain environments.

\section{Implementation and Evaluation} \label{imp_and_eva}
We implement a prototype of POLARIS and conduct a comprehensive evaluation covering system performance,  concurrent tests, comparisons analysis and key derivation optimization.

% \subsection{Test Tools and Experimental Setup}
We adopt Hyperledger Fabric\cite{fabric}, a high-throughput consortium blockchain platform, to implement decentralized identity management in POLARIS. Golang v1.22 is used for system development, and Redis serves as the Resource Server's database for low-latency, high-concurrency storage.

Experiments are conducted on a server equipped with dual Intel Xeon Silver 4310 CPUs (24 cores, 2.10GHz) and 128GB RAM. Time overhead and network performance are measured by logging requests and responses through a custom middleware. To evaluate system performance under different loads, we use wrk2\cite{wrk2}, a high-performance HTTP benchmarking tool, to assess throughput, latency, and stability.

\renewcommand{\theadfont}{\normalsize\bfseries} % 统一表头字体
\renewcommand{\theadgape}{\Gape[0pt][0pt]{}}   % 移除额外垂直间距
\renewcommand{\cellgape}{\Gape[0pt][0pt]{}}     % 移除额外垂直间距
\begin{table}[htbp]
    \centering
    \caption{Execution Time and Communication Overhead of Key Operations}
    \label{tab:performance}
    \renewcommand{\arraystretch}{1.4}
    \normalsize
    \resizebox{\columnwidth}{!}{%
        \begin{tabular}{|c|c|r|r|r|}
        \hline
        \multirow{2}{*}{\thead{Process}} & 
        \multirow{2}{*}{\thead{Operation}} & 
        \multirow{2}{*}{\thead{Exec Time\\(ms)}} & 
        \multicolumn{2}{c|}{\thead{Network (KB)}} \\
        \cline{4-5}
         & & & \thead{Send} & \thead{Receive} \\
        \hline
        \multirow{2}{*}{\thead{Registration}} 
          & Register DID                         & 2466     & 0.034 & 0.097 \\
          & Upload Resource                      & 58.981   & 0.932 & 0.131 \\
        \hline
        \multirow{2}{*}{\thead{Identity\\Management}} 
          & Request VC                           & 62.676   & 0.458 & 0.595 \\
          & Receive VC                           & 106.813  & 5.257 & 0.786 \\
        \hline
        \multirow{3}{*}{\thead{Access\\Control}} 
          & Authentication    & 62.188   & 0.307 & 5.781 \\
          & Authorization                        & 88.92    & 6.008 & 0.825 \\
          & Access Resource                      & 51.933   & 0.448 & 0.079 \\
        \hline
        \end{tabular}%
    }
\end{table}

\begin{figure*}[t]
\centering
\begin{subfigure}{0.48\textwidth}
    \centering
    \includegraphics[width=0.85\textwidth]{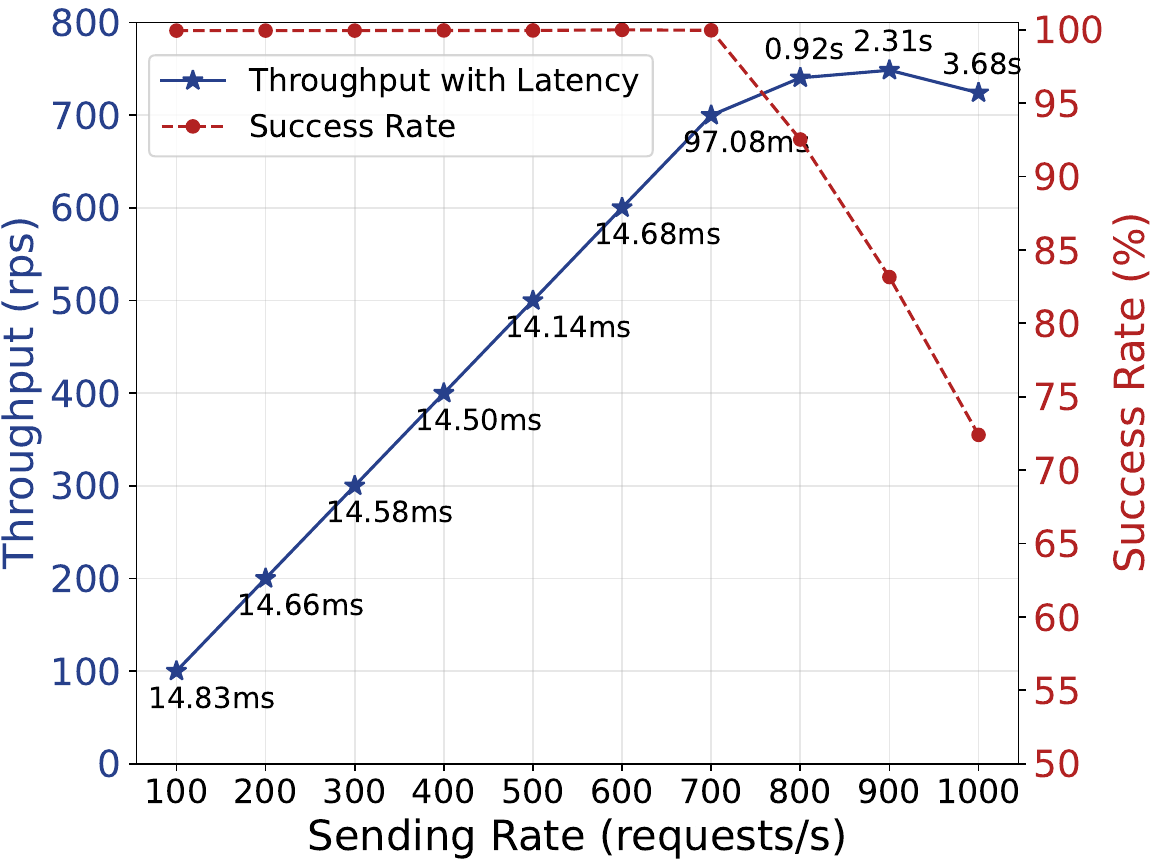}
    \caption{Read Throughput and Latency vs Sending Rate}
    \label{fig:workload_read}
\end{subfigure}%
\hfill
\begin{subfigure}{0.48\textwidth}
    \centering
    \includegraphics[width=0.85\textwidth]{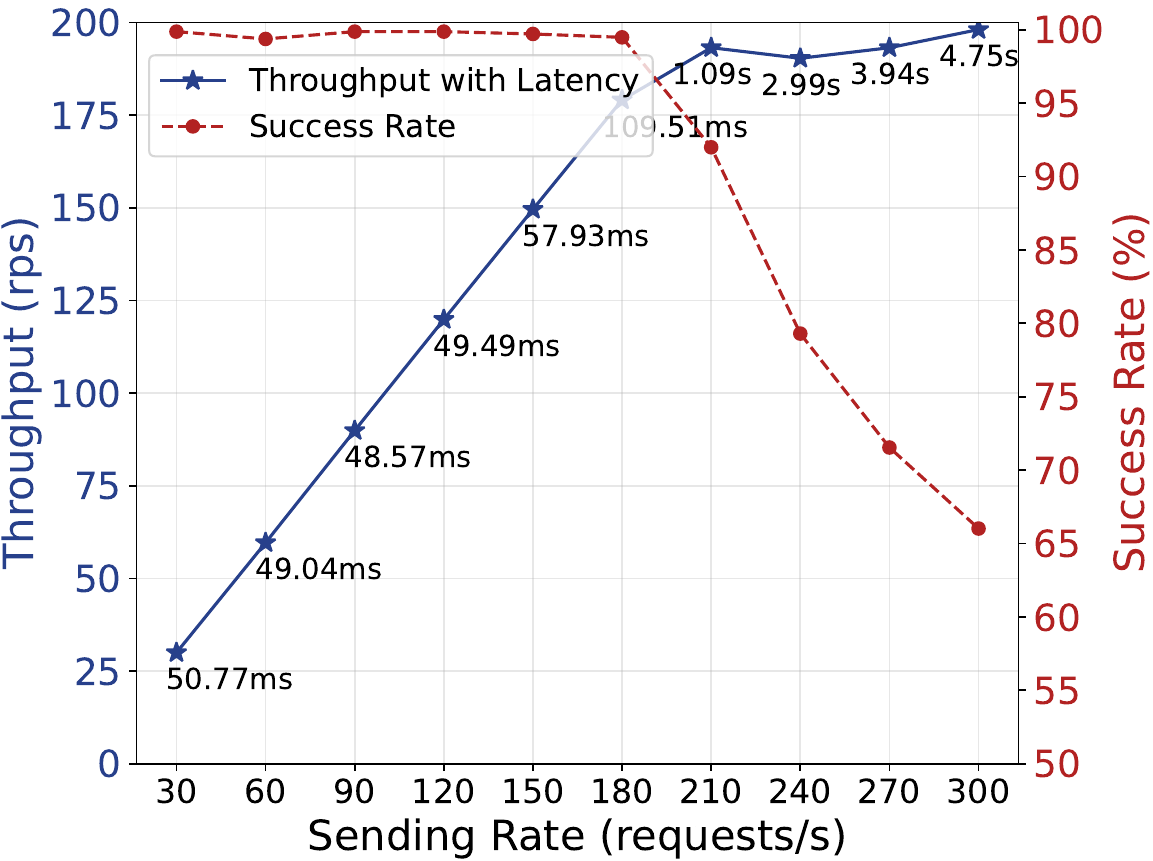}
    \caption{Write Throughput and Latency vs Sending Rate}
    \label{fig:workload_write}
\end{subfigure}
\caption{Throughput and latency involving read (Left) and write (Right) blockchain operations at different sending rates}
\label{fig:workload}
\end{figure*}

\begin{table*}[!t]
  \centering
  \scriptsize
  % 压缩列间距
  \setlength{\tabcolsep}{3pt}
  % 微调行高
  \renewcommand{\arraystretch}{1.1}
  \caption{Multi-Dimensional Comparison of POLARIS with Existing Works}
  \label{tab:comparison}
  \begin{tabular}{|
      >{\centering\arraybackslash}m{1.2cm}|
      >{\centering\arraybackslash}m{1.8cm}|
      *{11}{>{\centering\arraybackslash}p{1.2cm}|}
    }
    \hline
    \multirow{3}{*}{\shortstack[l]{\textbf{Category}}}
      & \multirow{3}{*}{\shortstack[l]{\textbf{Dimension}}}
      & \textbf{Lagutin, 2019\cite{Lagutin}}
      & \textbf{Jung, 2020\cite{Jung}}
      & \textbf{SSIBAC (Belchior, 2020) \cite{ssibac}}
      & \textbf{Kamboj, 2021 \cite{Kamboj}}
      & \textbf{Beomseok Kim, 2021\cite{Kim2021}}
      & \textbf{DSMAB (Saidi, 2022)\cite{Saidi}}
      & \textbf{Ma, 2022\cite{Ma}}
      & \textbf{Taehoon Kim, 2023 \cite{Kim2024}}
      & \textbf{Pandi- yarajan, 2024\cite{Pandiyarajan}}
      & \textbf{POLARIS (Ours)} \\
    \hline
    % 下面五行归属同一大类
    \multirow{5}{*}{\shortstack[l]{Architecture}}
      & SSI support
        & \checkmark & \checkmark & \checkmark & \xmark      & \checkmark & \checkmark & \checkmark & \checkmark & \checkmark & \checkmark \\ \cline{2-12}
      & DID support
        & \checkmark & \checkmark & \checkmark & \xmark      & \checkmark & \checkmark & \checkmark & \checkmark & \checkmark & \checkmark \\ \cline{2-12}
      & Blockchain-based
        & \checkmark & \xmark      & \checkmark & \checkmark  & \checkmark & \checkmark & \checkmark & \xmark      & \checkmark & \checkmark \\ \cline{2-12}
      & Cross-domain support
        & \checkmark & \checkmark & \checkmark & \xmark      & \checkmark & \checkmark & \checkmark & \xmark      & \checkmark & \checkmark \\ \cline{2-12}
      & Auditability / Traceability
        & \xmark      & \xmark      & \checkmark & \checkmark  & \checkmark & \checkmark & \checkmark & \checkmark & \checkmark & \checkmark \\
    \hline
    \multirow{7}{*}{\shortstack[l]{Access\\ Control}}
      & \multirow{3}{*}{\shortstack[l]{Authentication}}
        & OAuth2 with ACE extension  & DID and DPKI  & DID with VP and ZKP  & on-chain account password & DID with VP-based proof  & DID with public key  & DID with VP and ZKP   & double VC with CP-ABE  & DID with VP and ZKP  & DID with VP-based proof \\ \cline{2-12}
      & \multirow{2}{*}{\shortstack[l]{Model}}
        & OAuth2 & RBAC & ABAC with RBAC & RBAC & ABAC & ABAC & ABAC & ABAC with CP-ABE & ABAC & ABAC with issuer \\ \cline{2-12}
      & Credential Aggregation
        & / & / & \checkmark & / & \checkmark & / & \checkmark & / & / & \checkmark \\ 
    \hline  
    \multirow{9}{*}{\shortstack[l]{Policy}}
      & Policy Language
        & / & / & XACML & / & XACML & / & / & / & / & VPPL \\ \cline{2-12}
      & Policy Complexity
        & medium & low & high & low & medium & low & low & medium & high & high \\ \cline{2-12}
      & \multirow{2}{*}{\shortstack[c]{Policy Granularity}}
        & attribute level & role level & attribute level & role level & attribute level & data object level & attribute level & attribute level & attribute level & attribute level \\ \cline{2-12}
      & \multirow{4}{*}{\shortstack[c]{Policy Expression}}
        & condition-based conjunction & membership logic rules & context-aware privilege rules & role-only control logic & VC attribute binding & default attribute set & attribute list matching & tree-structured policy logic & attribute logic + Boolean ops & attribute-issuer logic composition \\
    \hline  
    \multirow{6}{*}{\shortstack[c]{Security\\ Features}}
      & Minimal Disclosure
        & \checkmark & \xmark & \checkmark & \xmark & \checkmark & \xmark & \checkmark & \checkmark & \checkmark & \checkmark \\ \cline{2-12}
      & User-controller Presentation
        & \checkmark & \checkmark & \checkmark & \xmark & \checkmark & \checkmark & \checkmark & \checkmark & \checkmark & \checkmark \\ \cline{2-12}
      & \multirow{3}{*}{\shortstack[c]{Session-level \\Security}}
        & Proof of Possession Token & one-time nonce mechanism & \multirow{3}{*}{/} & \multirow{3}{*}{/} & \multirow{3}{*}{/} & \multirow{3}{*}{/} & \multirow{3}{*}{/} & \multirow{3}{*}{/} & \multirow{3}{*}{/} & Key derivation mechanism \\
    \hline
  \end{tabular}
\end{table*}

\subsection{System Performance}\label{lab:performance_test} 
We conduct a systematic evaluation of POLARIS’s execution performance and communication overhead by decomposing the access control process into three staged workflows: \textbf{Registration}, \textbf{Identity Management}, and \textbf{Access Control}, each consisting of multiple atomic operations. The main results are summarized in Table~\ref{tab:performance}.

In terms of execution time, several operations within the identity management and access control processes (such as \textit{Request VC} and \textit{Authentication}) involve on-chain queries to the VDR, resulting in stable latencies about 62–106ms. The most time-consuming operation is the \textit{Register DID} operation in the registration process, taking around 2.47s due to blockchain write operations. However, since this operation is only triggered once during initial user onboarding and is not repeated in subsequent interactions, its one-time cost is considered acceptable.

In terms of communication overhead, atomic operations involving VC/VP transmission, such as \textit{Receive VC} and \textit{Authorization}, generate approximately 5.257KB and 6.008KB of network overhead respectively, which constitute the main bandwidth burden. This result is closely related to the size of the VC structure. The current test is based on a smaller VC template, real-world deployments with nested attributes or extended signature chains may introduce greater transmission overhead. To address this, POLARIS adopts a key derivation mechanism aligned with minimal disclosure principles (see Section~\ref{kd_opt}), which effectively reduces the amount of unnecessary data exchanged and improves communication efficiency.

In addition, resource-related operations such as \textit{Upload Resource} and \textit{Access Resource} demonstrate favourable performance, with execution times well below 60ms and network traffic under 1KB, indicating strong responsiveness and lightweight behaviour even in resource-intensive scenarios.

In summary, results show that POLARIS achieves verifiable access control processes with limited on-chain interaction and manageable communication overhead, demonstrating its efficiency and practical deployment potential.

\subsection{Concurrent Test}
To evaluate the scalability and stability of POLARIS under realistic load conditions, we conduct a series of concurrent performance tests targeting both read-intensive and write-intensive operations. These experiments are crucial, as real-world cross-domain identity interactions often involve large volumes of parallel requests, including credential queries, access control decisions, and identity registrations. Ensuring reliable performance under such stress is essential for practical deployment in dynamic, high-throughput environments.

In particular, we distinguish between two types of blockchain-related operations:

\begin{itemize}
    \item \textbf{Read-intensive operations:} such as Request VC and Authorization, reflect frequent identity verification requests across domains.
    \item \textbf{Write-intensive operations:} such as Register DID, simulate identity registration which are relatively heavier but less frequent.
\end{itemize}

The results are shown in Figure.~\ref{fig:workload}, with (a) illustrating read operations and (b) write operations. For read operations, POLARIS maintains near-linear throughput scaling up to 600 requests per second (rps), with latency stable under 10ms and a 100\% success rate, until hitting a saturation point around 800rps. For write operations, throughput scales linearly until around 180rps, after which latency rises and success rate begins to decline.

These tests cover the end-to-end access control workflow, including blockchain interactions, signature verification, and session-level security mechanisms, reflecting POLARIS’s responsiveness in typical cross-domain scenarios. The results show that POLARIS delivers robust concurrency performance across identity management and access control tasks, confirming its suitability for deployment in demanding environments.

\subsection{Comparative Analysis}

Unlike performance-critical components such as zero-knowledge proof circuits or blockchain consensus algorithms, decentralized access control frameworks currently lack standardized benchmarking environments or widely accepted evaluation datasets. Consequently, most existing works remain conceptual or prototype-level implementations with heterogeneous architectures and evaluation metrics, making direct quantitative comparison infeasible and potentially misleading. To ensure fairness, we therefore adopt a qualitative, multi-dimensional comparison aligned with POLARIS’s design objectives.

As summarized in Table~\ref{tab:comparison}, the evaluation spans four major aspects: \textbf{architecture}, \textbf{access control}, \textbf{policy}, and \textbf{security features}. Each aspect is examined along representative dimensions such as blockchain integration, cross-domain interoperability, model granularity, and session-level protection.

Compared with ZKP-heavy approaches such as BBS+-based VP, POLARIS adopts a structured commitment disclosure mechanism which achieves comparable privacy guarantees with significantly lower computational and deployment overhead. This lightweight design enables faster verification and better scalability for high-throughput scenarios, though it provides weaker formal unlinkability than zero-knowledge counterparts. Likewise, the use of a VDR built on blockchain enhances transparency but introduces latency and on-chain dependency; off-chain or hybrid registries may offer alternative trade-offs for IoT or latency-sensitive environments.

In summary, POLARIS provides comprehensive support across key dimensions of decentralized access control, achieving a balanced integration of architectural flexibility, fine-grained and verifiable policy logic, and privacy-preserving credential management. Its lightweight yet trustworthy design enables practical deployability at scale, while maintaining strong security and interoperability across diverse domains. Future work will further complement this analysis with empirical performance evaluations once compatible benchmarking environments are established.

\begin{figure}[h]
\centering
\includegraphics[width=0.48\textwidth]{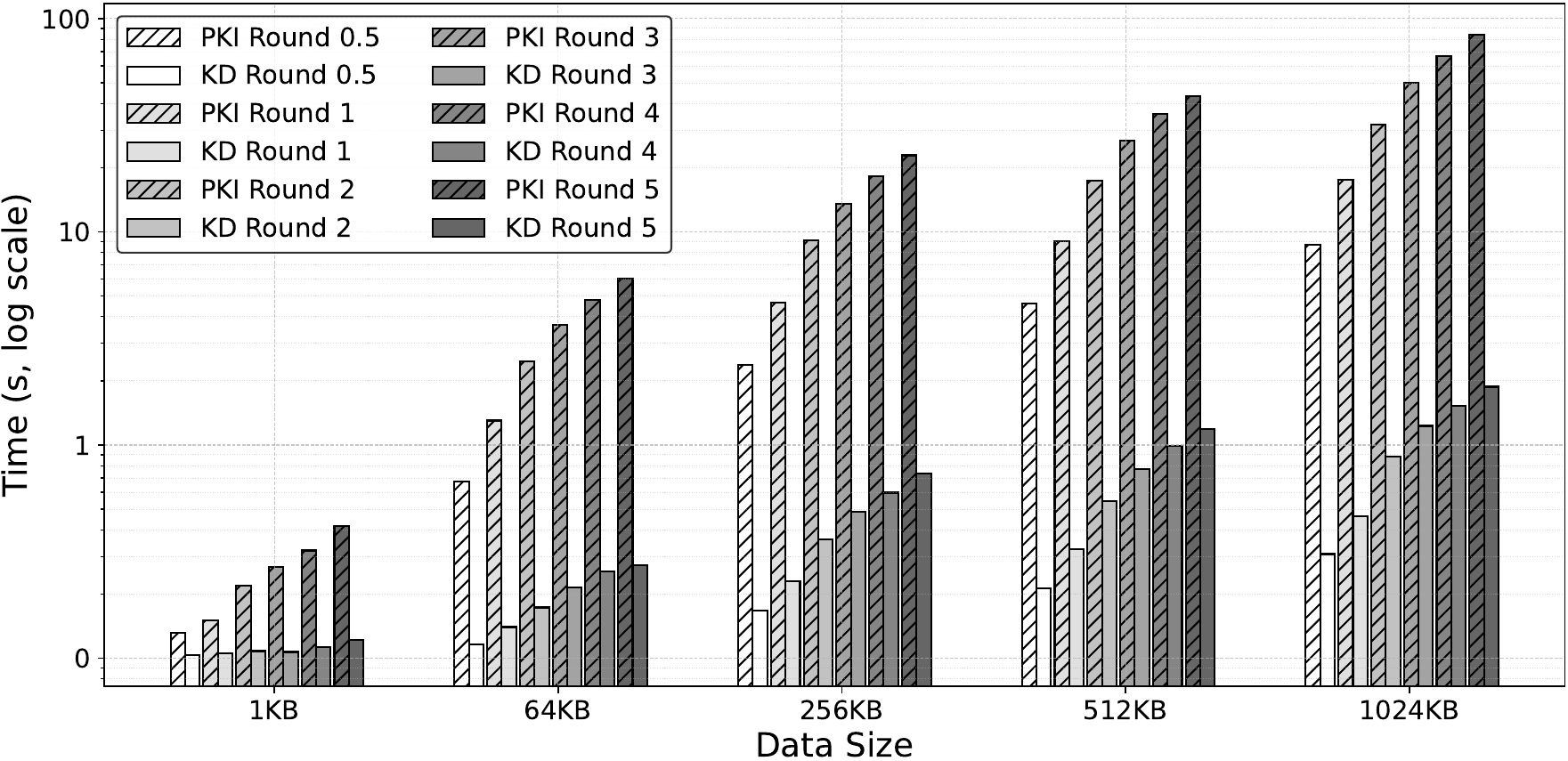}
\caption{Round vs Time for PKI and KD Methods.}
\label{fig:pki_vs_kd}
\vspace{-5pt}
\end{figure}

\subsection{Key Derivation Optimization}\label{kd_opt} 

In Section~\ref{lab:performance_test}, we mentioned that VCs and VPs can become significantly large in practical access control scenarios, especially when fine-grained policies demand comprehensive identity attributes. Under such conditions, ensuring secure and trustworthy transmission through encryption and digital signatures may introduce substantial time overhead.

To mitigate this challenge, we employ a session-level security mechanism based on the key derivation (KD) method introduced in Section~\ref{key_derivation}. This mechanism enables efficient and secure session key derivation between two communicating entities without repeated blockchain interactions. In addition to providing session isolation and flexibility in key length, the mechanism demonstrates notable advantages in reducing performance overhead, which we highlight in this section.

We evaluate and compare the encryption and signing latency under both the traditional PKI approach and the KD-based mechanism during access control exchanges. For PKI, we cache the retrieved public key for the entire session to avoid repeated blockchain access and simulate a fairer baseline. We vary the data sizes and the number of exchange rounds (where 0.5 round denotes a one-way message) to assess scalability.

As shown in Figure.~\ref{fig:pki_vs_kd}, the PKI overhead scales almost linearly with data size, with 1MB requiring up to 20 seconds in a single round. In contrast, the KD-based mechanism maintains consistently low overhead, staying below 2 seconds even with large-scale data and multi-round exchanges. These results demonstrate that our session-level security mechanism significantly reduces runtime cost and is well-suited for real-world, data-intensive access control environments.

\section{Conclusion} \label{conclusion}
In this paper, we propose \textbf{POLARIS}, a policy-based and privacy-preserving access control architecture designed for cross-domain environments. To address the challenges of trustworthy identity verification and fine-grained authorization across heterogeneous domains, POLARIS introduces three key mechanisms. First, the \textit{Structured Commitment Disclosure (SCD)} mechanism enables selective attribute disclosure with strong verifiability and integrity guarantees. Second, the \textit{Verifiable Presentation Policy Language (VPPL)} provides a lightweight yet expressive foundation for fine-grained, auditable, and multi-source policy evaluation. Third, a session-level security mechanism ensures binding between entity and resource access, enhancing session confidentiality and resistance to replay attacks.

Extensive evaluations demonstrate that POLARIS achieves secure, efficient and scalable access control in heterogeneous ecosystems. These results validate its potential as a foundational framework for future privacy-respecting and identity-aware systems that span organizational and trust boundaries.

\section*{Acknowledgment}
This work is funded by Fuxi Institution-CASICT Internet
Infrastructure Laboratory.

\bibliographystyle{IEEEtran}
\bibliography{references}

\end{document}